\documentclass[b5paper,twoside]{stm18b5}

\usepackage[pdftex]{graphicx} 

\usepackage[utf8]{inputenc}
\usepackage[T1]{fontenc}
\usepackage{euscript}
\usepackage{amsfonts,amssymb,amsmath}
\usepackage{amsthm}
\usepackage{enumerate}
\usepackage{amscd}

\usepackage[russian,english]{babel}
\usepackage{t1enc}

\usepackage{hyperref}

\def\url#1{\texttt{#1}}

\newcommand{\be}{\begin{equation}}
\newcommand{\ee}{\end{equation}}
\newcommand{\beq}{\begin{equation}}
\newcommand{\eeq}{\end{equation}}
\newcommand{\dee}[2]{\ensuremath{{{\rm d} #1\over {\rm d} #2}}}
\newcommand{\doo}[2]{\ensuremath{{\partial #1\over \partial #2}}}

\def\Div{{\rm div}}
\def\vg{{\bf g}}

\def\vk{{\bf k}}
\def\vq{{\bf q}}
\def\vx{{\bf x}}
\def\vv{{\bf v}}
\def\vf{{\bf f}}

\begin{document}

\setcounter{equation}{0}
\setcounter{figure}{0}
\setcounter{section}{0}


\thispagestyle{plain}
\addcontentsline{toc}{subsection}{\numberline{}\hspace*{-15mm}J.~Honkonen:
\emph{Randomly stirred perfect gas}}

\markboth{\sc J.~Honkonen} {\sc Randomly stirred perfect gas}

\STM

\title{Randomly stirred perfect gas}

\authors{Juha~Honkonen}

\address{Department of Military Technology, National Defence
University,\\ P.O.Box 7,00861, Helsinki, Finland}

\bigskip

\begin{abstract}
Foundations of the analysis of scaling in randomly stirred compressible fluid with the aid of stochastic differential equations are discussed in the example of perfect gas. The structure of the stress tensor with nonnegative shear and bulk viscosities is determined in $d$-dimensional space. It is argued that the steady cascade picture of energy transfer is compatible with generic hydrodynamic equations. A renormalizable model of randomly stirred polytropic fluid is put forward and it is shown that this model should be used for description of randomly stirred perfect gas instead of the model of ''isothermal'' fluid.
\end{abstract}

\vspace*{3mm}

\section{Introduction}

Since the pioneering work of Forster, Nelson and Stephen \cite{FNS76,FNS77}  as well as De Dominicis and Martin \cite{Dominicis79}
theoretical analysis of the problem of turbulence in incompressible fluid with the aid of functional methods and the renormalization group has grown to a significant branch of theoretical physics combining advanced methods of quantum field theory and classical stochastic problems \cite{turbo,Vasilev04,Antonov06,Hnatic16}.

The basic setup of the problem in the field-theoretic approach is the randomly forced Navier-Stokes equation together with the continuity equation:
\beq
\dee{\vv}{t}=\doo{\vv}{t} + \vv\cdot \nabla  \vv =\nu \nabla^2 \vv 
- {1\over \rho}\,\nabla p + \vf,
\quad  \dee{\log \rho}{t}=-\Div\, \vv
\label{eq:NS&CE}
\eeq
with the correlation function of the Gaussian random force $\vf$ (per mass unit)
\begin{equation}
d_f(k)=D_{1} k^{4-y-d}+D_{2} k^2\,,
\label{nakach2}
\end{equation}
Due to the condition of incompressibility 
\footnote{It is worth noting that the current basic reference of Landau and Lifshitz \cite{Landau_fluid} states this condition as the unconditional constant density $\rho$, whereas the classical monograph of Kochin {\em et al} \cite{Kochin63} has a slightly different point of view: $\dee{\rho}{t}=0$; both approaches yield the crucial condition
$\Div\,\vv=0$} the continuity equation reduces to the transversality condition to the velocity field $\Div\,\vv=0$. This feature leads to enormous simplification of the stochastic problem from the point of view of the complete set of hydrodynamic equations. First, the pressure is excluded from the transverse part of the Navier-Stokes equation, which becomes a closed equation for the divergenceless part of the velocity field. Second, the pressure is found as function of the velocity from the longitudinal part. Third, the continuity equation is exhausted by the transversality condition.
Therefore, in incompressible fluid the Navier-Stokes equation and the continuity equation are completely separated from the heat transfer equation, whose solution requires the knowledge of system-dependent equations of state of the fluid: the caloric equation of state for the specific internal energy $u=u(T,\rho)$ and the mechanical equation of state for the pressure $p=p(T,\rho)$. 

The solution of the stochastic problem (\ref{eq:NS&CE}), (\ref{nakach2}) is presented in a functional-integral form with the De Dominicis-Janssen action of the pair of the transverse velocity field $\vv$ and the transverse auxiliary vector field $\vv'$:
\begin{multline}
S[\vv,\vv'] = \frac{1}{2} \int\! dt\int\! d\vx \int\! d\vx'\mbox{ }v_i'(t,\vx)P_{ij}d_f(\vx-\vx')
v_j(t,\vx')\\
+ \int\! dt\int\! d\vx \mbox{ }v_i'(t,\vx)\left[-\doo{}{t} v_i(t,\vx)
+\nu \nabla^2 v_i(t,\vx) - v_j(t,\vx) \doo{}{x_j} v_i(t,\vx)
\right]\,.\nonumber
\end{multline}
In this presentation the advanced machinery of perturbative quantum field theory is then used for asymtotic analysis of the correlation functions of the random velocity field \cite{turbo,Vasilev04}.

Contrary to the incompressible case, relatively little has been done to develop a similar approach to randomly forced Navier-Stokes equation of compressible fluid, although turbulent transport in a compressible random field with given statistics has been an important topic in the RG analysis since the first successful results of its application to the Kraichnan-Obukhov model \cite{Antonov06}. Attempts to carry out a renormalization group (RG) analysis of stochastic hydrodynamics of compressible fluid have been based on
the idea to close the system of the Navier-Stokes equation and the continuity equation with assumption of ''isothermal'' fluid (i.e. a fluid, in which the density is a linear function of the pressure: $ (p - \overline{p}) = c^{2} (\rho - \overline{\rho})$, where $c$ is the speed of sound,) \cite{Staroselsky90,Antonov97,AntKos14,Antonov17,Honkonen18}.

There are, however, some issues in this approach. First, 
in order to preserve the homogeneity property of the free theory necessary for prescribing definite canonical dimensions to fields and parameters, an additional term must be introduced to the continuity equation which gives rise to the linear structure typical of diffusion equation. Although this term is brought about by renormalization and is thus self-consistent, it seems quite inappropriate from
the point of view of the classical hydrodynamic theory
\cite{Landau_fluid}. Second, the speed of sound appears in this approach as a ''mass'' parameter. The asymptotic behaviour predicted by the scaling dimensions calculated with the aid of the RG holds, strictly speaking, when all mass parameters are put equal to zero. Therefore, predictions of the model about the asymptotic behavior apply to the case of zero speed of sound, i.e. to ultimate hypersonic turbulence. 

For most practical purposes a compressible fluid is perfect gas and in this report the equation of state of the perfect gas $p=\rho T$ and the specific internal energy $u=c_VT$ will be used whenever the thermodynamic properties of the compressible fluid are needed. Having explicit equations of state means that the number of independent variables in hydrodynamic equations is just equal to the number of equations. Therefore, any additional restriction may turn out to be incompatible with the standard hydrodynamic equations and must be checked with respect to this.

This paper is organized as follows. In section \ref{sec:HDEqs} the complete set of hydrodynamic equations with random stirring is recapitulated in arbitrary space dimension $d$ and the structure of the viscous stress tensor is analyzed. Section \ref{sec:barotropic} is devoted to analysis of the consistency of the energy cascade picture from the point of view of the complete set of hydrodynamic equations. In Section \ref{sec:polytropic} an analysis of renormalization of the randomly stirred polytropic perfect gas is carried out.

\section{Hydrodynamic equations in $d$ dimensions}
\label{sec:HDEqs}

Hydrodynamic equations are conservation laws. In compressible fluid with transsonic characteristic velocities shock waves appear and the differential equations of hydrodynamics are not valid everywhere. In stochastic problems, however, there are -- as a rule -- no smooth solutions anyway. Therefore, in the problem of randomly stirred fluid the stochastic differential equations (SDE) actually hold in a much wider class of functions than in the deterministic problem. However, even this case solutions to the SDE are usually continuous functions although non-differentiable, which at least formally excludes shock waves as solutions. On the other hand, if the characteristic velocity of the fluid flow is supersonic or even hypersonic it might be expected that the contribution of subsonic and transsonic modes is negligible. In order to keep things simple and consider a homogeneous gas of uncharged particles without radiation effects, the temperature of the gas must be low. Nonwithstanding provisions to the applicability of stochastic hydrodynamic equations to physical problems, the stochastic problem is interesting in itself and will be discussed as such.

In the RG approach formulation of the stochastic problem is almost invariably given in arbitrary space dimension $d$, therefore the hydrodynamic equations of compressible fluid will be recapitulated here for that case.
The momentum equation gives rise the Navier-Stokes equation for the velocity field $\vv$
\be
\rho\dee{v_i}{t}=\doo{\sigma_{ij}}{x_j}+\rho f_i\,,\nonumber
\ee
where $\rho$ is the (mass) density although henceforth number density will be assumed, $\sigma_{ij}$ is the stress tensor and $f_i$ the bulk force per mass unit. The stress tensor that includes the effect of forces at the boundaries of the fluid volume. Apart from electromagnetically active fluids, the only bulk force is the gravity, which can hardly be random in Earthly conditions.
The effect of any realistic random stirring thus should be described through the random part of the stress tensor. 
However, it is necessary to use random bulk forcing to describe the cascade picture of energy transfer. The mechanical power input by the bulk force is the source of energy in the energy balance equation (which is not actually a conservation law), but it is absent in the heat transfer equation. The stress tensor appears in both equations with opposite signs of the mean energy transfer. In order to describe consistently a steady state for the
energy balance equation, it is assumed that the energy input at large scales is transferred by the eddy cascade to small scales and dissipated by viscous terms. What is not always said is that the energy dissipation in energy balance equation is the source of energy in the heat transfer equation and by virtue of this equation causes accumulation of internal energy, which is not a stationary process.

The stress tensor in three dimensions is
\be
\label{3stress}
\sigma_{ij}=-\delta_{ij}p+\eta\left(\doo{v_i}{x_j}+\doo{v_j}{x_i}-{2\over 3}\delta_{ij}\doo{v_l}{x_l}\right)+\zeta\delta_{ij}\doo{v_l}{x_l}\,,
\ee
where $\eta$ and $\zeta$ are the dynamic shear and bulk viscosities, respectively. In (\ref{3stress})
the irreducible part is explicitly written as a separate term and contains the space dimension three in the coefficient of the Kronecker symbol. In $d$ dimensions the coefficient is thus $2/d$ instead of $2/3$ used -- quite surprisingly -- in earlier work \cite{Staroselsky90,Antonov97,AntKos14,Antonov17,Honkonen18}. This is reflected in the connection between the dynamic viscosities $\eta$ and $\zeta$ on one hand and kinematic viscosities $\mu$ and $\nu$ on the other. It should be emphasized that this is not a mere choice of notation. As will be shown below, both viscosities appear in the heat transfer equation as well and the positivity of dynamic viscosities follows from the second law of thermodynamics when the shear viscosity $\eta$ is defined as the coefficient of the irreducible part of the stress tensor.  Thus, in $d$ dimensions
the stress tensor should be written in the form
\be
\label{d-stress}
\sigma_{ij}=-\delta_{ij}p+\sigma'_{ij}=-\delta_{ij}p+\eta\left(\doo{v_i}{x_j}+\doo{v_j}{x_i}-{2\over d}\delta_{ij}\doo{v_l}{x_l}\right)+\zeta\delta_{ij}\doo{v_l}{x_l}\,,
\ee
with positive viscosities $\eta$ and $\zeta$. Here, $\sigma'_{ij}$ is the viscous stress tensor. In case of explicitly position-independent dynamic viscosities the forced Navier-Stokes equation is of the form
\beq
\label{NSE}
\dee{v_{i}}{t} = - {1\over \rho}\doo{p}{x_i}+\nu \left(\delta_{ij}\nabla^{2} - \doo{^2}{x_i\partial x_j}\right) v_{j} 
+ \mu \doo{^2}{x_i\partial x_j} v_{j}  + f_{i}\,,
\eeq
where the kinematic viscosities are (note the $d$ dependence)
\beq
\nu={\eta\over \rho}\,,\quad \mu={1\over \rho}\left(\zeta+{2d-2\over d}\eta\right)\,.
\nonumber
\eeq
What has not been discussed at all in the previous RG analyses of the compressible fluid with fluctuating forces is the total energy conservation the focus having been on the energy balance equation. With the standard set of physically consistent external sources of energy at moderate temperatures (to exclude radiative energy transfer) the energy conservation law in hydrodynamics is
\beq
\label{EnergyConservation}
\rho\left({1\over 2}\,\dee{v^2}{t}+ \dee{u}{t}\right)=\kappa\nabla^2 T+\rho\,\vf\cdot\vv+\doo{}{x_i}\,\sigma_{ij}v_j\,,
\eeq
where $u$ is the specific internal energy and $\kappa$ the heat conductivity. 
The heat transfer equation follows from the conservation of energy with the account of mechanical work and heat conduction \cite{Kochin63}. For the present purposes it is convenient to write the heat transfer equation in terms of internal energy.
Substitution in (\ref{EnergyConservation}) of the kinetic energy from the energy balance equation obtained by scalar multiplication of the Navier-Stokes equation by the velocity field yields the heat transfer equation
\begin{multline}
\label{heatEq}
\rho \dee{u}{t}=\kappa\nabla^2 T+\sigma_{ij}\doo{v_j}{x_i}
=
\kappa\nabla^2 T\\
-p\,{\rm div}\vv+{\eta\over 2}{\rm Tr}\left(\doo{v_i}{x_j}+\doo{v_j}{x_i}-{2\over d}\delta_{ij}\doo{v_l}{x_l}\right)^2+\zeta \left({\rm div}\vv\right)^2\,.
\end{multline}
and we see that the power of the bulk force is absent in the heat transfer equation. In order to express the viscous stress tensor part on the right side as a sum of two squares it is necessary to define the viscosities in the $d$-dimensional case according to (\ref{d-stress}).

Contrary to momentum equations and heat transfer equation, the continuity equation has no adjustable coefficients:
\be
\dee{\rho}{t}=-\rho\,{\rm div}\,\vv\,.\nonumber
\ee
On one hand this makes it look the same for both incompressible and compressible fluid, on the other there is nothing to be renormalized. It is hard to imagine that in continuum case the density would be scale dependent.

According to the Landau-Lifshits theory of hydrodynamic fluctuations \cite{Landau_stat2}
thermal fluctuations are introduced in hydrodynamic equations as normally distributed
random viscous stress tensor $s_{ij}$ and random heat flux density $\vg$, i.e.
\beq
\sigma_{ij}\to \sigma_{ij}+s_{ij}; \vq=-\kappa\nabla T\to -\kappa\nabla T+\vg\,.\nonumber
\eeq
When these random sources are included in the
the energy conservation law (\ref{EnergyConservation}), it is immediately seen that in a translation invariant system the bulk force $\vf$ is the only possible source of mean energy injection to the system, contributions of the heat flux and stress tensor vanish on averaging. The energy balance equation obtained from the Navier-Stokes equation by scalar multiplication by the velocity field allows to conclude that in case of normally distributed random bulk force the energy injection to the fluid due to the random force is positive. This means that in case of random stirring by random bulk force -- which is rather unphysical -- from energy conservation law (\ref{EnergyConservation}) it follows that the full energy of the fluid (integral of the left side of (\ref{EnergyConservation}) over the fluid volume) is a monotonically growing function of time. Thus, in this case no steady state in the fluid is not obvious even in the case of incompressible fluid, contrary to the implicit assumption of the vast majority of work on the randomly stirred fluid. Richardson's cascade of energy transfer from large to small scales used for a phenomenological description then is tantamount to an additional assumption that the whole energy input goes to the increase of the internal energy. 

We recall that in modelling of turbulence a random source of momentum and energy in the from of a normally distributed random force is introduced to maintain a steady state of the turbulent flow. An invariable tacit assumption in this construction is that the statistically homogeneous random forcing generates a steady state in the flow. What has been shown above is that for a physically consistent randomly stirred flow this may not be true: a normally distributed random bulk force does not maintain a steady state in the whole. On the contrary, it leads to monotonically growing energy in the fluid.
In case of incompressible flow the assumption of the accumulation of all additional energy in the internal energy saves the day and the steady state is maintained for the mechanical degrees of freedom. It turns out that the same conclusion holds for all barotropic fluid flows, as will be shown in the next section.

\section{Randomly stirred barotropic fluid}
\label{sec:barotropic}

In a barotropic fluid the density is a function of the pressure only: $\rho=\rho(p)$. From this property and the thermodynamic relation for the specific internal energy
\[
du=c_VdT+\left[p-T\left(\doo{p}{T}\right)_\rho\right]\,{d\rho\over \rho^2}
\]
it immediately follows that
\[
du=c_VdT+{p(\rho)\over \rho^2}d\rho\,.
\]
This representation implies that the specific heat $c_V$ is independent of the density and the internal energy of a barotropic fluid 
is a sum of a function of the temperature and a function of the density
\begin{equation}
u=\int c_V(T)dT+\int {p(\rho)\over \rho^2}d\rho=u_1(T)+u_2(\rho)\,.\nonumber
\end{equation}
Specific enthalphy -- introduced as the Legendre transform of internal energy -- obeys a similar decomposition
\[
h=u+{p\over \rho}=h_1(T)+h_2(p)\,.
\]
In barotropic fluid $c_V=c_p$, therefore $h_1=u_1$ and the function $h_2$ is the Legendre transform of $u_2$:
\begin{equation}
h_2(p)=\int {p(\rho)\over \rho^2}d\rho+{p\over \rho}\,,\qquad dh_2={dp\over \rho}\,.\nonumber
\end{equation}
The expression of the differential of $h_2$ is exactly of the same form as the pressure gradient term in the Navier-Stokes equation (\ref{NSE}). It is therefore convenient to introduce a new variable (to simplify notation henceforth, a new symbol is introduded)
\beq
\label{def phi}
\phi\equiv h_2\,,\qquad d\phi={dp\over \rho}\,.
\eeq
In these terms the Navier-Stokes equation (\ref{NSE}) becomes
\beq
\label{NSE-phi}
\dee{v_{i}}{t} = - \doo{\phi}{x_i}+\nu \left(\delta_{ij}\nabla^{2} - \doo{^2}{x_i\partial x_j}\right) v_{j} 
+ \mu \doo{^2}{x_i\partial x_j} v_{j}  + f_{i}\,,
\eeq
To obtain a closed system of equations for $\vv$ and $\phi$, calculate the time derivative of $\phi$, which with the use of the continuity equation yields
\beq
\dee{\phi}{t}=-\doo{p}{\rho}\Div\, \vv\,.\nonumber
\eeq
In practical applications the most popular barotropic models are the incompressible fluid ($\rho= {\rm const}$), ''isothermal'' fluid
$p=C\rho$ and polytropic fluid $p=C\rho^{n}$, where $n$ is the polytropic index. In case of isothermal fluid we have
\[
\dee{\phi}{t}=-{ C}\Div\, \vv\,,
\]
which leads to the model used in the earlier RG analyses of compressible fluid \cite{Antonov97,AntKos14,Antonov17,Honkonen18}. In the next section it will be shown that this is not the model suitable for description of perfect gas.
In polytropic fluid the time derivative becomes a nonlinear function of the fields
\beq
\label{Dt-phi-poly}
\dee{\phi}{t}=-(n-1)\phi\,\Div\, \vv\,.
\eeq
This changes the perturbation theory drastically. It is also the case well suited for the description of randomly forced perfect gas. Since there is no explicit dependence on the temperature in the Navier-Stokes equation (\ref{NSE-phi}) and the evolution equation (\ref{Dt-phi-poly}), these equations constitute a closed system of equations for $\vv$ and $\phi$.

The density-dependent part of the internal energy may be incorporated in the energy balance equation, which assumes the form
\beq
\label{EnergyBalanceBarotropic}
\rho\left({1\over 2}\,\dee{v^2}{t}+ {p\over \rho^2}\dee{\rho}{t}\right)=
-\doo{}{x_i}pv_i+
\rho\,\vf\cdot\vv+v_j\doo{}{x_i}\,\sigma'_{ij}\,.
\eeq
The total energy function of \cite{Galtier11} is a particular case (isothermal) of the functions of left side of (\ref{EnergyBalanceBarotropic}).
The generic form of the barotropic specific energy function (i.e. the temperature-independent part of the total specific energy) assumed to be stationary in the cascade picture is
\beq
\label{mech-en}
e'={1\over 2}v^2+\int {p(\rho)\over \rho^2}d\rho\,.
\eeq
In the ''isothermal'' fluid 
\beq
\label{e'isothermal}
e'={1\over 2}v^2+C\log {\rho\over \rho_0}\,,
\eeq
whereas in the polytropic fluid
\beq
\label{e'polytropic}
e'={1\over 2}v^2+{C\over n-1}\rho^{n-1}\,.
\eeq
In (\ref{e'isothermal})
the internal energy is assumed to vanish at the reference value the density $\rho=\rho_0$, while in (\ref{e'polytropic}) the internal energy vanishes with the density.

As a consistency check calculate the time derivative of the function $u_2$
\beq
\dee{u_2}{t}=-{p\over \rho}\Div\, \vv\nonumber
\eeq
and substitute in the heat transfer equation (\ref{heatEq}) upon which the
contribution of
the function $u_2$ is cancelled. Thus, the
heat transfer equation (\ref{heatEq}) becomes an equation for the dynamics of the temperature of the form
\begin{equation}
\label{heatEqBarotropic}
\rho \dee{u}{t}=\rho c_V\dee{T}{t}
=
\kappa\nabla^2 T
+{\eta\over 2}{\rm Tr}\left(\doo{v_i}{x_j}+\doo{v_j}{x_i}-{2\over d}\delta_{ij}\doo{v_l}{x_l}\right)^2+\zeta \left({\rm div}\vv\right)^2
\end{equation}
lacking any explicit dependence on the pressure $p$ and thus the field $\phi$.
Assuming the cascade transfer of the energy from the work of the random force to heat through viscous terms of the Navier-Stokes equation (\ref{NSE-phi}) and the heat transfer equation (\ref{heatEqBarotropic}) (they sum up to a total derivative and thus on the average are equal and of opposite sign).
It is seen that the result of the energy injection is the change ot the temperature which does not affect the ''mechanical'' energy (\ref{mech-en}) of the barotropic fluid, for which a steady state may be assumed on the same basis as in the incompressible case.

\section{Polytropic perfect gas. Power counting}
\label{sec:polytropic}

In the perfect gas the linear dependence of the density on the pressure means that the fluid is genuinely isothermal. In the perfect gas the internal energy -- which is a function of temperature only -- in this case is a constant and in the cascade picture for the energy flux coming from the random forcing there is nowhere to go. Therefore, the model of ''isothermal'' fluid is not applicable to the perfect gas. It is worth noting that this is the case always, when the barotropic fluid is at a constant temperature. According to the  heat transfer equation of the barotropic fluid (\ref{heatEqBarotropic}) there is no sink for the energy transferred through the cascade in this case either.

In case of barotropic perfect gas, i.e. a polytropic gas the internal energy is not fixed and it is conceivable that it will absorb the injected energy via increase of the temperature. Therefore it seems consistent to adopt the cascade picture and the closure of hydrodynamic equations in terms of the Navier-Stokes equation (\ref{NSE-phi}) and the evolution equation for the scalar field (\ref{Dt-phi-poly}). Standard procedures give rise to the basic action \cite{Vasilev04} (in a shorthand notation with all integrals and sums implied)
\begin{multline}
\label{BasicA1}
S[\vv,\vv',\phi,\phi'] = \frac{1}{2}v_{i}' D_{ij} v_{j}' + 
v_{i}' \Biggl[- \dee{}{t} v_{i}
+\nu \left(\delta_{ij} \nabla^{2} - \doo{^2}{x_i\partial x_j}\right) v_{j}  
+ u \nu \partial_{i} \partial_{j} v_{j}\\ - \doo{}{x_i} \phi \Biggr]
-\phi' \left[ \doo{}{t} \phi+(1+\lambda_1)v_i\doo{}{x_i}\phi  \right]-\lambda_1\phi v_i\doo{}{x_i}\phi'\,.
\end{multline}
Power counting for one-irreducible Green functions of this field theory reveals that a couple relevant terms are generated by renormalization which are not present in (\ref{BasicA1}). The model is logarithmic, when the space dimension $d=4$ and the falloff parameter $y=0$ in the force correlation function (\ref{nakach2}). This immediately lends the term proportional to $k^2$ irrelevant in the correlation function. The degree of divergence of a one-irreducible graph of the logarithmic model is
\beq
\delta=6-N_v-3N_{v'}-2N_\phi-2N_{\phi'}\,,\nonumber
\eeq
where $N_i$ is the number of arguments of the graph corresponding to field $i$.
In the analysis of possible generation terms -- on top of the arguments used in earlier work \cite{Antonov97,Antonov17} -- it should be taken into account that the propagator and vertex structure of the perturbation expansion due to (\ref{BasicA1}) is such that in one-irreducible Green functions the number of $\phi$ arguments $N_\phi$ is no less than the number of $\phi'$ arguments $N_{\phi'}$. The result of inspection of graphs is that on top of field monomials already present in basic action (\ref{BasicA1}), the following relevant terms are generated by renormalization and added to the basic action to arrive at a multiplicatively renormalizable model:
\beq
\phi'\nabla^2\phi\,,\qquad \phi'\phi^2\,.\nonumber
\eeq
Thus, the basic action of the multiplicatively renormalizable model may be written as
\begin{multline}
\label{BasicAMulti}
S[\vv,\vv',\phi,\phi'] = \frac{1}{2}v_{i}' D_{ij} v_{j}' + 
v_{i}' \Biggl[- \dee{}{t} v_{i}
+\nu \left(\delta_{ij} \nabla^{2} - \doo{^2}{x_i\partial x_j}\right) v_{j}  
 + u \nu \partial_{i} \partial_{j} v_{j}\\ - \doo{}{x_i} \phi \Biggr]
 -\phi' \left[ \doo{}{t} \phi+(1+\lambda_1)v_i\doo{}{x_i}\phi - \chi\nu \partial^{2} \phi +{\lambda_2\over 2}\phi^2\right]-\lambda_1\phi v_i\doo{}{x_i}\phi'\,.
\end{multline}
Due to Galilei invariance, there is only one coupling constant $\lambda_1$ in the two drift terms of the scalar field.

With the account of the generation term
the force correlation function is
\beq
D_{ij}(\vk) = g_{1} \nu^{3} k^{4-d-y}\left[P_{ij}(\vk) + \alpha Q_{ij}(\vk)\right] +
g_{2} \nu^{3}\,,\nonumber
\eeq
where $ g_{1}$ and $g_{2} $ are coupling constants and $ P_{ij}(\vk) = \delta_{ij} - k_{i}k_{j}/k^{2} $ 
and $ Q_{ij}(\vk) = k_{i}k_{j}/k^{2} $ are transversal and longitudinal projection operators.

Although the vector-field part of the action (\ref{BasicAMulti}) is the same as in the ''isothermal'' case, there are important differences in the scalar part of the action. First, the physical content of the scalar field is different corresponding to (\ref{e'polytropic}) instead of (\ref{e'isothermal}). Moreover, the scalar field $\phi$ is defined by the mechanical degrees of freedom of the internal energy according to (\ref{def phi}) and is therefore bound to have phenomenological parameters which may well be renormalized.  Thus, there no physical provisions to renormalization of the evolution equation of $\phi$ contrary to that of the density $\rho$.
Second, there is no ''mass'' term in (\ref{BasicAMulti}), therefore the Green functions produced automatically obey powerlike asymtotic behaviour defined by scaling exponents. Third -- and this is rather rare in turbulent transport problems -- the time derivative of the scalar field is renormalized due to the appearance of the nonlinear term in on the right side of the evolution equation (\ref{Dt-phi-poly}). Fourth, a new nonlinear term is generated by renormalization.

\end{document}